\documentclass{aastex}          %% The manuscript based on AASTeX v5.x
\usepackage{spr-astr-addons}    %% mimicing ASTR journal style
\usepackage{color}
\usepackage{textcomp}

\begin{document}
%% Article title
%
\title{Double eclipsing binary system KIC 3832716}

%% Running heads
\shorttitle{Double eclipsing binary system KIC 3832716}
\shortauthors{Fedurco et al.}

%% Author and Affilations
\author{M. Fedurco\altaffilmark{1}} \and \author{\v{S}. Parimucha\altaffilmark{1}}
%\and 
%\author{\altaffilmark{}}
%\affil{}
\email{miroslav.fedurco@student.upjs.sk} %% non-output

%% Alternate Affilations
\altaffiltext{1}{Institute of Physics, Faculty of Science, Pavol Jozef \v{S}af\'{a}rik University, Park Angelinum 9, 040 01, Ko\v{s}ice, Slovakia}
%\altaffiltext{2}{}
%\altaffiltext{3}{}

%% Abstract
\begin{abstract}
In this paper we analyzed the light curve of the object KIC~3832716 observed during Kepler K1 mission. We showed that this previously known eclipsing binary is in fact quadruple system, in double eclipsing binary configuration on a long period orbit with the mass ratio of eclipsing binaries $0.7 \pm 0.3$. The system consists of eclipsing binary A with the orbital period of $P_A \sim 1.1419$\,d. Eclipsing binary A contains larger but less luminous secondary component in the post main sequence stage of its evolution consistent with an "Algol paradox" which can explain parameters of the components by mass transfer from secondary component after leaving the main sequence. Inspection of the light curve and eclipse time variations of the eclipsing binary A also indicates the presence of spots on the surface of secondary component with $57$\,d period of activity probably induced by longitudonal motion of the spots. The second eclipsing binary B with orbital period $P_B \sim 2.1703$\,d contains very dim secondary component with luminosity below 0.09\,$L_{\odot}$.
\end{abstract}

%% Keywords

\graphicspath{{figs/}}
\section{Introduction}
The Kepler satellite, launched in 2009, has produced observations with unprecedented photometric precision \citep{Borucki}. It has revolutionized the study
of extrasolar planets, variable stars and stellar astrophysics; providing photometric data with high-precision, high-cadence continuous light curves.  After losing two reaction wheels, the Kepler spacecraft ended its primary mission and started its so-called K2 mission \citep{Howell2014}. The photometric precision of K2 is slightly lower, but still much better than that from ground-based observatories.

The analysis of excellent scientific data from Kepler can reveal not only extrasolar planets and new variable stars \citep{molnar16}, but detailed inspection of light curves of known systems can uncover various physical processes, such as pulsations of component(s) in eclipsing binary and/or presence of other bodies in the systems \citep{Gies, Zasche}. An example of such a system is also KIC~3832716, the analysis of which is presented in this paper.

\begin{table}[t]
\small
\caption{Basic information about KIC 3832716 (ep=J2000, eq=2000)}
\label{tab:dataspecs}
\begin{tabular}{c c c}
\tableline
Parameter & Value & Source \\
\tableline
$RA$ & $19^h\,01^m\, 34.6^s$ & \citet{Skrutskie06} \\
$DE$ & $38^{\circ}\, 54'\, 17.69''$ & \citet{Skrutskie06} \\
$Kp/[mag]$ & $13.42$ & \citet{Conroy} \\
$T_{eff1}/[K]\tablenotemark{a}$ & $6500\pm400$ & \citet{Armstrong14} \\
$\log(g)(cgs)$ & 4.15 & Kepler MAST \\
$Q_{SC}\tablenotemark{b} $ & 2,4,5 & \\
$N_{SC}\tablenotemark{c}$ & 383\,532 & \\
$Q_{LC}\tablenotemark{b}$ & 0-17 & \\
$N_{LC}\tablenotemark{c}$ & 65\,307 \\
d/[kpc]\tablenotemark{d} & 2.1(2) & \citet{Lindegren18}\\
\tableline
\end{tabular}
\tablenotetext{a}{effective temperature of the primary component}
\tablenotetext{b}{quarters of observational data used for each cadence}
\tablenotetext{c}{total number of data-points analyzed for given cadence}
\tablenotetext{d}{distance to object (Gaia)}
\end{table}

Multiplicity of KIC~3832716 was detected during a survey for planetary candidates conducted by \citet{Borucki11}. After automatic detection of transit-like variations, the object was flagged as a false positive due to stellar nature of the eclipses. The object was included in the first release of Kepler eclipsing binary catalog \citep{Prsa11} as Algol-type eclipsing binary star with the orbital period of $\sim$1.142\,d. In the second data release of Kepler eclipsing binary catalog \citep{Slawson11} another ephemeris with the period of $\sim$2.170\,d was discovered. According to the catalog, the nature of the second set of eclipses was also stellar but without any further specification. The object was not listed in the list of blended sources and contamination factor of $0.030$  suggests that the light curve suffers minimally from a background light. In this paper we showed that the second ephemeris corresponds to the second eclipsing binary (EB). Additionally, \citet{Davenport16} detected multiple flares in the light curve using automated procedure which we confirmed by a visual inspection of the observations.

From this point forward, for the sake of clarity, we will refer to the central eclipsing binary with orbital period $\sim$1.142\,d as EB~A and to the second eclipsing binary with orbital period $\sim$2.170\,d as EB~B.  
In this paper we provide times of minima and ephemerides for both sets of eclipses using the method based on fitting of template curves (Sections~\ref{sec:EBAephemeris} and \ref{sec:res_crv}) together with a method to separate both sets of eclipses (Section~\ref{sec:phs_smth}). Photometric parameters of EB~A are derived in Section~\ref{sec:phb}. In Section~\ref{sec:ETTV} we examine eclipse time variations and a possible explanation of these changes is given in Section~\ref{sec:spots}. Our results are discussed in Section~\ref{sec:disscusion}.

\begin{figure}[t]
 \includegraphics[width=\columnwidth]{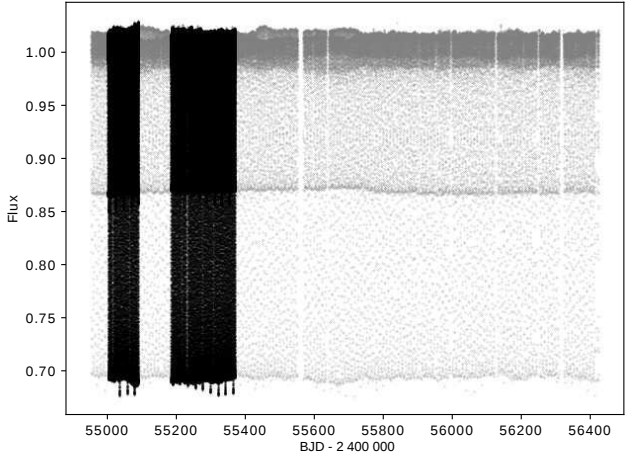}
 % light_curve.eps: 0x0 pixel, 300dpi, 0.00x0.00 cm, bb=18 180 594 612
 \caption{De-trended light curve of KIC~3832716 obtained in short cadence (black points) and long cadence regimes (gray points).}
 \label{fig:light_crv}
\end{figure} 

\section{Observational data}
For our analysis we used de-trended (PDCSAP\_FLUX) data from the third revision of the Kepler eclipsing binaries catalog \citep{Conroy}. 
We used short cadence (SC) data (sampled every 58.8\,s) from quarters Q2, Q4 and Q5 and long cadence (LC) data (sampled every 29.4\,min) from quarters Q1 to Q17 \citep{Gilliland}. KIC~3832716 was observed approximately 261\,d in SC regime and 1334\,d in LC regime. Basic information about the object and observational data is listed in Tab. \ref{tab:dataspecs}. De-trended light curves observed in both regimes are displayed in Fig. \ref{fig:light_crv}.

The initial visual inspection of the light curve of KIC 3832716 shows that it is typical for a detached EB with noticeable ellipsoidal variations between eclipses.
Eclipses caused by the second eclipsing binary EB~B are also visible in the original light curve during coincidences between eclipses caused by both binaries as periodic variations in depth of EB~A eclipses. However, much larger variations in flux caused by EB A are preventing us from analyzing the eclipses of EB~B directly from the original light curve and thus pre-whitening of the EB~A signature becomes necessary (see sec. \ref{sec:res_crv}).

\begin{figure}[t]
 \includegraphics[width=\columnwidth]{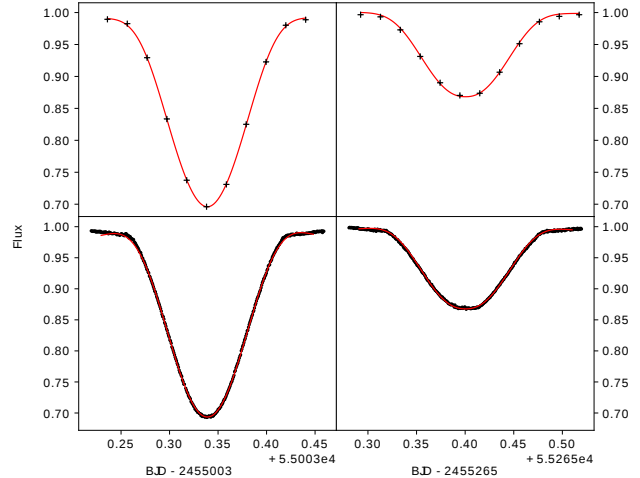}
 \caption{Primary (left column) and secondary eclipses (right column) of the EB A with their template eclipse functions. Eclipses observed in LC or SC are displayed separately in top and bottom row. Template eclipse functions are displayed as red lines.}
 \label{fig:minimum_fit}
\end{figure}

\section{Ephemerides of eclipsing binaries}
\label{sec:ephemerides}

\subsection{Ephemeris of EB A}
\label{sec:EBAephemeris}
Preliminary orbital period $\sim$1.14188\,d of EB~A was calculated using the Phase Dispersion Minimization (PDM) method \citep{Stellingwerf}. This value together with the time of the first observed primary eclipse was used to calculate the positions of all other observed eclipses. The exact times of minima were determined by fitting a template eclipse function to observed eclipses.  We used the template eclipse function proposed by \citet{mikulasek} that was originally intended for phenomenological modeling of eclipsing binary light curves. Template eclipse function (eq. 13 in \citet{mikulasek}) contains 6 free parameters for each type of eclipse (primary and secondary) and they were determined by the fitting of phase stacked eclipses. We decided to fit only the light curve around eclipses in phase interval from 0.88 to 1.12 in case of primary eclipses, and from 0.38 to 0.62 in case of secondary eclipses. In order to reduce the number of fitted parameters, we performed fitting separately for both types of eclipses. We also derived different template eclipse functions for eclipses observed in LC and SC regimes. 

Preliminary values of fitted parameters were obtained using genetic algorithms. Subsequently, Markov chain Monte-Carlo (MCMC) simulation with $10^6$ iterations was used to obtain statistically significant values of parameters. 

\begin{figure}[t]
  \includegraphics[width=\columnwidth,keepaspectratio=true]{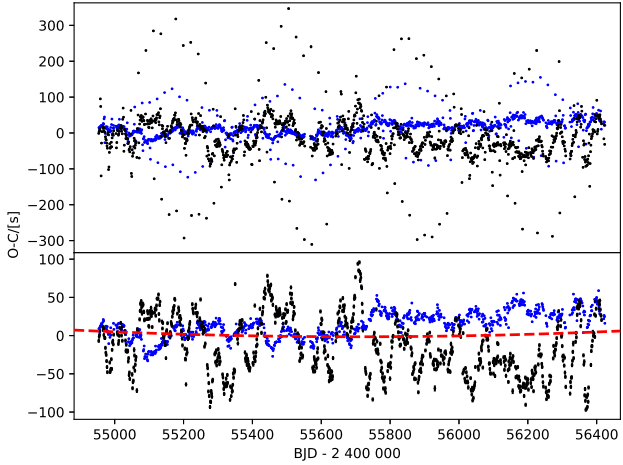}
  \caption{O-C diagram for primary (black points) and secondary eclipses (blue points) of EB A. Top panel shows original O-C data with periodic spikes caused by coincidences of central EB A eclipses with eclipses caused by EB B. Bottom panel shows quadratic fit to the O-C data with removed spikes as the red dashed line.}
  \label{fig:OC}
\end{figure}

Afterwards, template eclipse functions were used to fit particular eclipses. In this case, we fixed all parameters except parameter $A$ which represents depth of an eclipse and the phase shift parameter $\Delta\vartheta$. $\Delta\vartheta$ was introduced to the model by substituting phase parameter $\vartheta$ with $\vartheta -\Delta\vartheta$. This enabled us to shift the position of template eclipse function by the amount of $\Delta\vartheta$ and finally to determine the time of minimum for a particular eclipse. MCMC simulation with $10^6$ iterations was used to find parameters $A$ and $\Delta\vartheta$ for each eclipse and to properly estimate their standard deviations. Especially important was the standard deviation of the parameter $\Delta\vartheta$ from which the error in time of minima can be derived. Examples of primary and secondary eclipses of EB A observed in both regimes with their template eclipse functions are displayed in Fig. \ref{fig:minimum_fit}. 

In the case of SC observations, we were able to obtain times of minima with standard deviation of approximately $1.2$\,seconds for primary and $2.5$\,seconds for secondary eclipses. Surprisingly, almost the same precision was achieved for times of minima of eclipses observed in LC regime. Although LC data have 30 times longer sampling period, much higher precision of LC observations and sufficient width of eclipses ($\sim$3.5\,hours) were able to produce times of minima with almost the same precision for both regimes.

All determined times of minima were used to improve the orbital period obtained from the Kepler eclipsing binaries catalog and provide a linear ephemeris of EB~A. O-C diagram of primary and secondary minima constructed using this orbital period is displayed in the top panel of Fig. \ref{fig:OC}. 

O-C data of the EB A shows very complex structure which is the result of multiple effects. The most prominent feature is the presence of periodic spikes. The height of the spikes changes with the period of 350\,d and the spikes themselves occur with the period of about 22\,d. 

\begin{figure}[t]
  \includegraphics[width=\columnwidth,keepaspectratio=true]{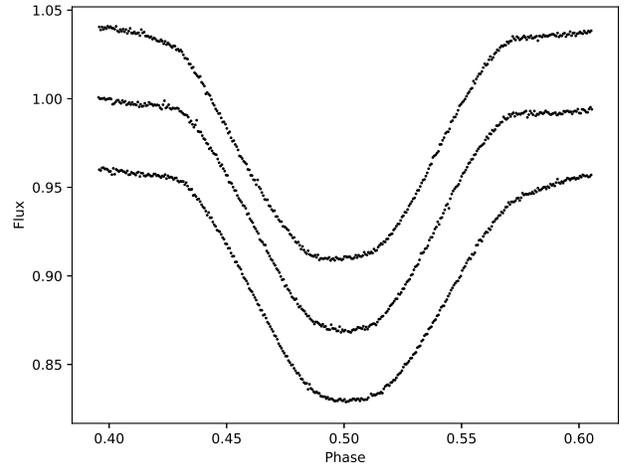}
  \caption{An example of the deformation of eclipses due to the presence of EB B eclipse during the secondary eclipse of EB A. Top curve shows an example where eclipse from EB B precedes the secondary eclipse of EB A and bottom curve shows the situation where EB B eclipse follows the secondary EB A eclipse. As a reference, middle curve shows unaffected secondary EB A eclipse.}
  \label{fig:deformation}
\end{figure}

These spikes are produced only when the eclipse of EB~B is coinciding with the primary or secondary eclipse of EB~A. Mutual overlap of EB A and EB B eclipses deforms overall shape of the light curve and therefore the time of minimum of EB A eclipse is shifted towards the position of nearby EB B eclipse as illustrated in Fig. \ref{fig:deformation}.

The ratio of orbital periods of both binaries $P_A$ and $P_B$ (Section~\ref{sec:res_crv}) is close to 1:2, thus the period of spikes can be calculated assuming this resonance of orbital periods:

\begin{equation}
 P_s=\frac{P_AP_B}{2P_A-P_B}\approx21.8\,d,
\end{equation}
and the observed period of amplitude variation of spikes can be explained by 10:1 resonance between the period of spikes $P_s$ and orbital period $P_B$: 
\begin{equation}
 P_m=\frac{P_BP_s}{P_s-10P_B}\approx350\,d.
\end{equation}

Knowing the nature of the spikes, we can now remove them and investigate short term variations of the O-C diagram. Analysis of these short term variations is discussed in detail in Sections \ref{sec:ETTV} and \ref{sec:spots}. We also performed quadratic fit to the O-C data without spikes to examine any long term non-linear effects. The resulting quadratic fit is displayed in the bottom panel of Fig. \ref{fig:OC}. We detected a slight increase in the observed orbital period of EB~A with the rate of $\dot{P}_A = dP_{A}/dt \approx 0.011(4)$\,s/yr. Therefore, for the final ephemeris of the EB~A we can write:

\begin{equation}
 \begin{split}
 Min~I = & ~2455055.86515(1)+1.14187687(3)\times E \\ 
         & + 2.0(7)\times10^{-10} \times E^2. 
 \end{split}
 \label{eq:ephem}
\end{equation}

\begin{figure}[t]
  \includegraphics[width=\columnwidth,keepaspectratio=true]{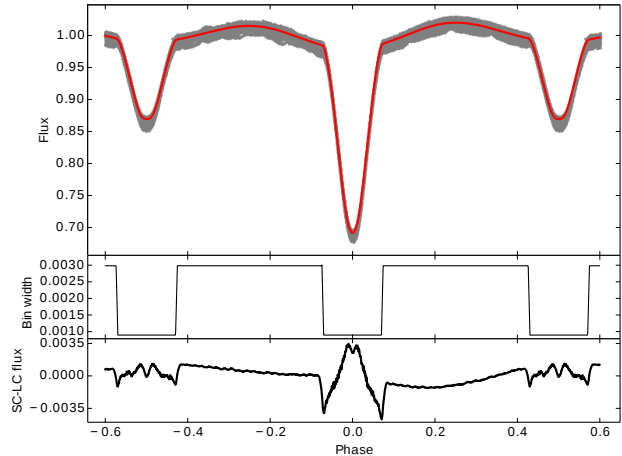}
  \caption{Top panel shows the smoothed phase curve (red line) of EB A described in detail in section \ref{sec:phs_smth} and original phase curve (gray dots). Middle panel shows the width of bins from which the smoothed phase curve was calculated. Comparison between smoothed phase curves calculated from both cadences is displayed in the bottom panel which clearly shows presence of the smearing effect occurring mainly during eclipses due to the longer sampling periods of the LC observations.}
  \label{fig:smooth_crv}
\end{figure}

\begin{figure}[t]
 \includegraphics[width=\columnwidth,keepaspectratio=true]{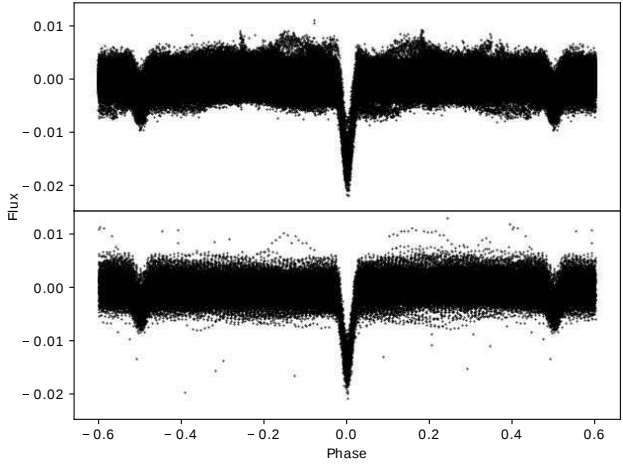}
  \caption{Phase curves of the EB B created from residual curve using SC (top panel) and LC data (bottom panel). Linear part of ephemeris in eq. \ref{eq:ephem2} was used to create both phase curves.}
  \label{fig:phs_trans}
\end{figure}

\subsection{Phase curve smoothing and residual curve}
\label{sec:phs_smth}
Further analysis of the EB A requires us to significantly reduce the number of data points in the phase curve. Also, in order to study the second light curve of EB B we needed to create a residual curve which would be free from the binary signature of EB~A. Both problems can be solved by using the smoothed phase curve of EB A. Smoothed phase curve was used to produce residual light curve by subtracting it from the original light curve and it was also used directly for the light curve analysis of EB A (section \ref{sec:phb}) due to the lower number of data points.

Initially, we divided the phase curve to $N_b = P_{A}/T_{SC}$ bins, where $T_{SC} = 58.8$\,s is the sampling period of SC observations. Afterwards, points of the smoothed phase curve along with their uncertainties were generated as weighted averages of flux for data in each bin. Despite lower sampling frequency, the same number of bins was also used for LC observations because of the sufficient number of evenly distributed data points across the phase curve. Problem emerging for the bins at the edges of the phase curve was solved by introducing cyclic boundary conditions, where the last data bin is immediate neighbor to the first data bin.

Due to different slopes of the phase curve during eclipses and outside of eclipses, we used different bin widths for each part of the phase curve as shown in Fig. \ref{fig:smooth_crv} (middle). Phase width of the bins was set to 0.001 during eclipses and to 0.003 outside of eclipses. Linear transitions between two bin widths were made at the edges of eclipses to prevent the creation of artifacts. The resulting smoothed phase curve of EB~A is shown in Fig.~\ref{fig:smooth_crv} (top). We tried several different widths of bins for both SC and LC regimes in order to find the optimal values for them. We found, however, that the above mentioned values of bin widths were able to smooth out all of the fluctuations of flux in the resulting smoothed phase curve below the precision of observations and simultaneously introduced the least amount of additional smearing into the data (see  Fig.~\ref{fig:smooth_crv}, bottom).

\begin{figure}[t]
 \includegraphics[width=\columnwidth,keepaspectratio=true]{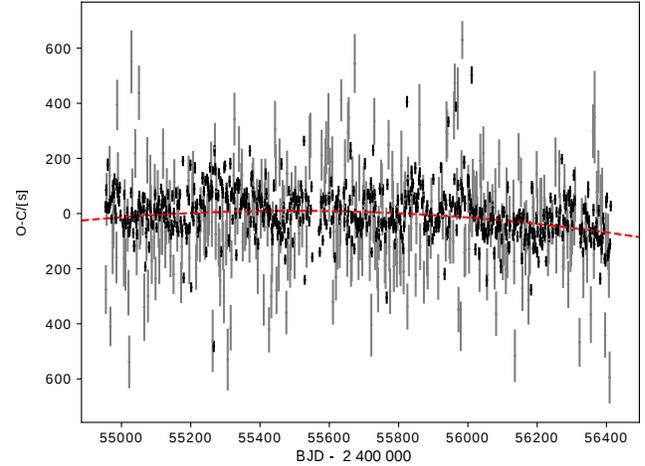}
  \caption{O-C diagram of EB~B, created using primary (black points) and secondary minima (gray points). Quadratic fit of O-C data is displayed as red dashed line.}
  \label{fig:OC_transit}
\end{figure}

\subsection{Ephemeris of the EB B}
\label{sec:res_crv}
Smoothed light curve of EB~A was subtracted from the original Kepler light curve (for LC and SC regimes separately) to obtain the residual curve, which is actually light curve of the second binary EB~B. These residual curves for SC and LC observations are almost free from flux variations caused by EB A as one can see in the phase curve of EB B in Fig. \ref{fig:phs_trans}. Using the approach of smoothed light curve  enabled us to recover even eclipses which coincide with EB~A eclipses and use them for the determination of the times of minima. 

Minima times of EB~B were calculated by a similar procedure as those of EB~A (Section \ref{sec:EBAephemeris}). One additional step has been taken before template fitting. Due to variable spot activity on the surface of EB A component (see Section~\ref{sec:spots}), smoothed phase curve was not able to completely remove the signature of EB A from the residual curve. Therefore, an additional detrending of each eclipse was performed by a quadratic fit. 
We reached the precision of 20\,s for primary eclipses and 90\,s for secondary eclipses. The resulting quadratic ephemeris of the second EB B was determined:

\begin{equation}
 \begin{split}
  Min~I_{B} = & 2455003.9077(2) + 2.1702736(3)\times E \\ 
	   & - 5(1)\times10^{-9} \times E^2. 
 \end{split}
 \label{eq:ephem2}
\end{equation}
O-C diagram of EB~B is shown in Fig. \ref{fig:OC_transit}, where shortening of the orbital period $P_B$ with the rate of $\dot{P}_B \approx -0.15(4)$\,s/yr is visible.

\begin{table}[t]
\small
 \caption{Photometric solution of EB~A binary.}  
 \label{tab:phb}
 \begin{tabular}{l c c}
 \tableline
 Parameter & Primary (1) & Secondary (2) \\
           & component  & component  \\
 \tableline
 $T(K)$ & 6500 & 5550(350) \\
 % $\Omega$ & 5.76(3) & 3.56(2) \\
 $r_{polar}\tablenotemark{a}$ & 0.20(2) & 0.26(1) \\
 $l_i\tablenotemark{b} $ & 0.48(8) & 0.41(6) \\
 %$R_2/R_1$ & \multicolumn{2}{c}{1.51} \\
 $i(^{\circ})$ & \multicolumn{2}{c}{78(1)} \\
 %$q$ & \multicolumn{2}{c}{0.6(1)} \\
 $l_{B}$\tablenotemark{c} & \multicolumn{2}{c}{0.12(9)} \\
  \tableline
 \end{tabular}
 \tablenotetext{a}{values of radii are listed in semi-major axis (SMA) units}
 \tablenotetext{b}{relative luminosities of the primary and secondary component \\ of  EB A}
 \tablenotetext{c}{total relative luminosity of EB B system, \\ where $l_1 + l_2 + l_{B} = 1$}
\end{table}

\section{Parameters of EB~A binary system}
\label{sec:phb}
Smoothed phase curve calculated in section \ref{sec:phs_smth} is now suitable for the light curve analysis of EB~A binary system. For this purpose we used software PHOEBE 0.31a \citep{Prsa_Zwitter}. The orbital period and shape of the phase curve strongly suggests a system with circular orbit, therefore we fixed the value of eccentricity to $e=0$. Effective temperature of the primary component was fixed to value $T_{1}=6500 \pm 400\,K$ \citep{Armstrong14}. We decided to adopt this value instead of value 5926\,K listed in Kepler Input Catalog (KIC) mainly due to the fact that KIC value is based on a single star model, whereas value published by \citet{Armstrong14} was calculated  considering both stars in the eclipsing binary using color information from multiple photometric surveys. We also fixed coefficients of gravity darkening $g_1$ = $g_2$ = 0.32 and the bolometric albedo coefficients $A_1$ = $A_2$ = 0.6, which are appropriate values for the convective envelopes \citep{Prsa}. The \cite{Kurucz93} model of stellar atmospheres was applied to the stars assuming solar composition. The limb darkening coefficients were interpolated from the van Hammes tables \citep{vanHamme} using linear cosine law. 

\begin{figure}[t]
  \includegraphics[width=\columnwidth,keepaspectratio=true]{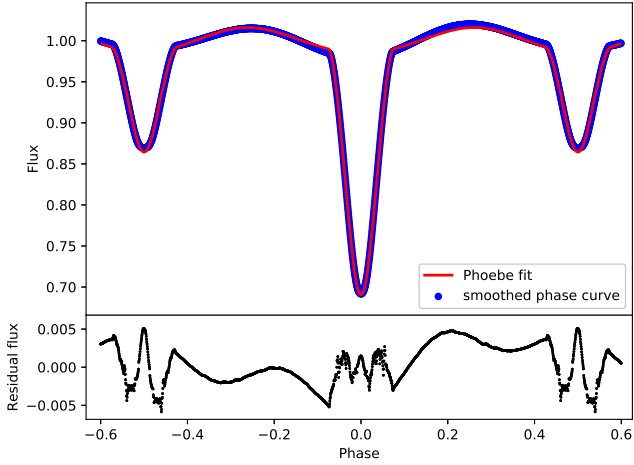}
  \caption{PHOEBE fit of EB A plotted in red against the smoothed phase curve (blue) used as input data. Noteworthy are artifacts during eclipses caused by inadequate treatment of limb darkening and surface discretization which led us to use smoothed phase curve instead of synthetic curve from PHOEBE to create residual light curve.}
  \label{fig:fit_res}
\end{figure}

In our solutions, we were looking for the temperature of the secondary component $T_2$, surface potentials of both of the components $\Omega_1$, $\Omega_2$, orbital inclination $i$ and the photometric mass ratio $Q$. We also searched for the contribution of the second binary EB~B in the system to total flux as the third light parameter $l_{B}$. The resulting parameters are listed in Table \ref{tab:phb} and the best fit is shown in Fig.~\ref{fig:fit_res}. 

\begin{figure}[t]
  \includegraphics[width=\columnwidth,keepaspectratio=true]{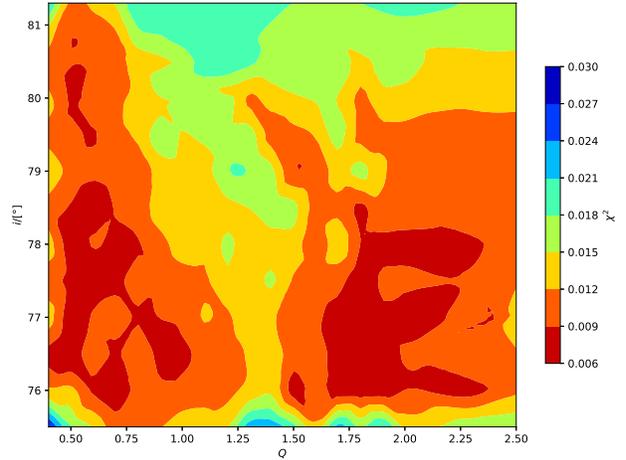}
  \caption{Quality of the light curve fit as a function of mass ratio $Q$ and inclination $i$. Binary systems models with acceptable quality of the fit ($\chi^2< 0.010$) are divided into two major groups. The first group for solutions with mass ratio below 1.1 and the second set of solutions for mass ratio above 1.5.}
  \label{fig:qi_trial}
\end{figure}

Due to possible correlation between mass ratio $Q$ and inclination $i$ we decided to perform an extensive search for other suitable solutions in $Qi$ plane around initial solution found via PHOEBE's differential corrections method. During this search we fixed values of $Q$ and $i$ and the rest of the parameters were fitted in similar fashion as in the case of the initial solution. We searched the $Q$ parameter space in range from 0.4 to 2.5 and $i$ parameter space in range from 75.5\,$^{\circ}$ to 81.3\,$^{\circ}$. The resulting $\chi^2$ distribution is displayed in Fig. \ref{fig:qi_trial}. As depicted in the Fig. \ref{fig:qi_trial} we were unable to determine photometric mass ratio of the system. We suspect that the main reason behind this uncertainty is the fact that the amount of third light contributed by EB B was unknown prior fit and it varied widely in our solutions across $Qi$ plane. This creates large uncertainty in absolute amplitude of ellipsoidal variations which is crucial in deriving photometric mass ratio. 

Despite the previously mentioned issues, statistical analysis of models with quality of the fit $\chi^2< 0.010$ provided us with very consistent values of effective temperature for the secondary component, polar radii and relative luminosities across the searched parameter space $Qi$. Final parameters of the binary system are listed in Table \ref{tab:phb}.   

Detailed inspection revealed that out of eclipse parts of the phase curve are slightly deformed with different maxima in flux. The most probable cause of this effect is the presence of spot(s) on the surface of the secondary component. Such deformation also decreases the accuracy of photometric mass ratio, determined by our solution. 

Our model indicates that EB~A is currently a detached binary with tidally deformed secondary component. The radius ratio of 1.30 indicates a larger secondary component. This accompanied by lower effective temperature than the primary component suggests that we are probably seeing an evolved, post main sequence (MS) secondary component.

We decided to test if our set of effective temperatures for primary and secondary component is consistent with effective temperature provided by KIC by calculating approximate aggregate effective temperature of both components. Resulting value of 5960\,K is in agreement with KIC value of 5926\,K and value from Gaia archive 5821\,K \citep{Sartoretti18}.

\section{Parameters of EB B}
\label{sec:EBBparams}
Despite large uncertainty in the light contribution of the EB~B to the overall flux of the system we attempted to derive its parameters. It was assumed that all of the third light detected in Section \ref{sec:phb} originated from the EB B and there are no other sources of light present in the system. We decided to investigate cases with different light contributions of the EB B based on the third light $l_{B}$ (Tab. \ref{tab:phb}) within its uncertainty to assess the reliability of EB~B parameters. The residual light curve after prewhitening of the EB A signature (see fig. \ref{fig:phs_trans}) was  renormalized using corresponding value of the third light and the output was analyzed in PHOEBE with no other third light. Resulting parameters of the EB B are listed in Tab. \ref{tab:EBBparams}. Parameters such as polar radii and relative luminosities suggest a system with much larger and luminous primary component accompanied by darker secondary component that contributes only 4\,\textperthousand~ to the total flux of the quadruple system. Due to the same reason as in the case of EB A, uncertainty in $l_B$ precludes the determination of photometric mass ratio and it is presumably responsible for the lack of precision of inclination as well. Additionally, because of the lack of any color information about EB B, only ratio of effective temperatures is listed in Tab. \ref{tab:EBBparams}.       

\renewcommand{\arraystretch}{1.3}
\begin{table}[t]
\small
 \caption{Photometric solution of EB~B binary.}  
 \label{tab:EBBparams}
 \begin{tabular}{l c c}
 \tableline
 Parameter & Primary (3) & Secondary (4) \\
           & component  & component  \\
 \tableline
 $T_{eff4}/T_{eff3}$ & \multicolumn{2}{c}{$0.70_{-0.12}^{+0.04}$} \\
 % $\Omega$ & 5.76(3) & 3.56(2) \\
 $r_{polar}\tablenotemark{a}$ & $0.13_{-0.03}^{+0.02}$ & $0.052_{-0.003}^{+0.016}$ \\
 $l_i\tablenotemark{b} $ & $0.964_{-0.014}^{+0.003}$ &$ 0.036_{-0.003}^{+0.014}$ \\
 %$R_2/R_1$ & \multicolumn{2}{c}{1.51} \\
 $i(^{\circ})$ & \multicolumn{2}{c}{$84_{-2}^{+4}$} \\
 %$q$ & \multicolumn{2}{c}{0.6(1)} \\
  \tableline
 \end{tabular}
 \tablenotetext{a}{values of radii are listed in SMA units}
 \tablenotetext{b}{relative luminosities of the primary and secondary component of \\ EB B, where $l_1 + l_2 + l_{B}(l_3 + l_4) = 1$}
\end{table}

\begin{figure}[t]
  \includegraphics[width=\columnwidth,keepaspectratio=true]{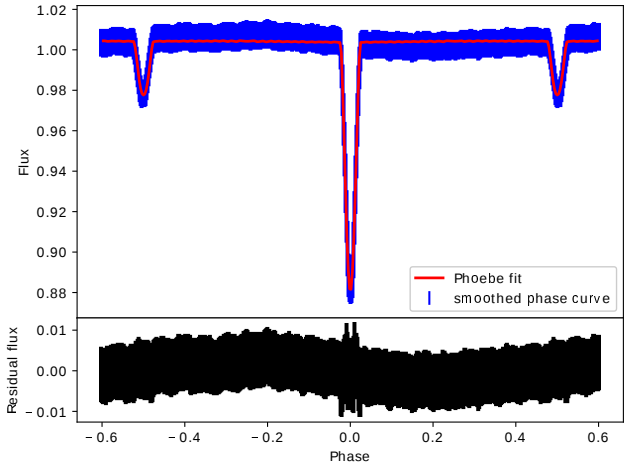}
  \caption{PHOEBE fit of EB B plotted in red against the renormalized smoothed phase curve (blue) using $l_B=0.12$ as the light contribution of EB B to the overall flux.}
  \label{fig:phb_fit_transit}
\end{figure}

\section{Luminosity estimation based on Gaia data}
\label{sec:luminosity}

We estimated total luminosity of the system using distance obtained from Gaia (see Tab.~ \ref{tab:dataspecs}) and coordinate dependent band extinction parameters $A_\lambda$ from \cite{Schlafly11}. Estimates were done in 2MASS J,H,K bands. Resulting luminosities ranged from 12.8 to 13.4\,L$_\odot$ with average uncertainty 2.4\,L$_\odot$. 

Using 13\,L$_\odot$ as an approximate total luminosity of the system and the relative luminosities calculated during light curve analysis we calculated estimated luminosities for all components of the quadruple system for 3 different values of the third light based on value $l_B$ from Table \ref{tab:phb} and its uncertainty in order to investigate the effect of uncertainty in $l_B$ on luminosities of the quadruple system members (see Tab. \ref{tab:luminosities}).

\begin{figure}[t]
 \includegraphics[width=\columnwidth,keepaspectratio=true]{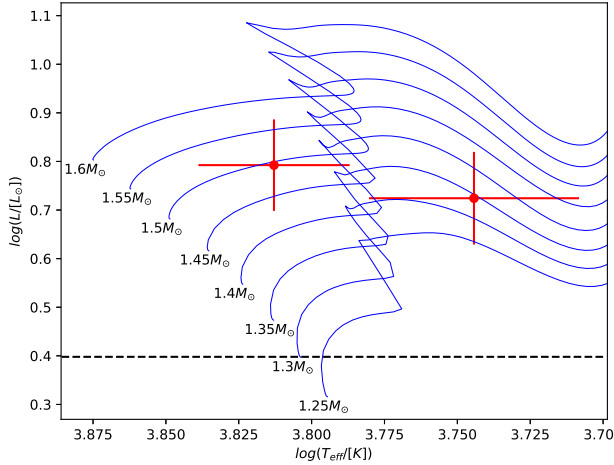}
  \caption{HR diagram with positions of the components of EB A system along with stellar evolutionary tracks modeled in MESA. Evolutionary models were calculated for masses from 1.25 to 1.6\,$M_{\odot}$ with assumed metallicity Z = 0.02 and initial hydrogen abundance $X_{ini}$ = 0.7. Ledoux criterion for convection was used along with overshooting parameter $f_{ov}=0.007$ above hydrogen burning core. Horizontal dashed line indicates upper luminosity estimate of primary component of EB B which is probably still on the MS due to much lower luminosity compared to components of EB A.} 
  \label{fig:HR}
\end{figure}

Luminosity and effective temperature of the primary component of EB A is consistent with MS star with mass around 1.5\,$M_{\odot}$ based on simulations carried out in Modules for Experiments in Stellar Astrophysics - MESA \citep{Paxton18}. However, luminosity of the secondary component of EB A is much too high for given effective temperature which puts it among evolved stars above MS as illustrated in Hertzprung-Russell (HR) diagram in Fig. \ref{fig:HR}. Uncertainty in the light contribution of the EB B prevents us from drawing any meaningful conclusions about its primary component. However, in case of the secondary component, we can conclude that obtained luminosity range is consistent with the expected luminosity range of a red dwarf. 

\begin{table}[t]
\small
 \caption{Luminosity estimation of the components based on total luminosity calculated from Gaia and relative luminosities calculated in tables \ref{tab:phb} and \ref{tab:EBBparams}.}
 \label{tab:luminosities}
 \begin{tabular}{l c c c c}
 \tableline
 & \multicolumn{2}{c}{EB A} & \multicolumn{2}{c}{EB B} \\
 $l_B$ & $L_1(L_{\odot})$ & $L_2(L_{\odot})$ & $L_3(L_{\odot})$ & $L_4(L_{\odot})$ \\
 \tableline
 0.03 & 6.8 & 5.8 & 0.4 & 0.01 \\
 0.12 & 6.2 & 5.3 & 1.4 & 0.05 \\
 0.20 & 5.6 & 4.8 & 2.5 & 0.09 \\
 \tableline
 \end{tabular}
\end{table}

\section{Eclipse time variations}
\label{sec:ETTV}
\begin{figure}[t]
 \includegraphics[width=\columnwidth,keepaspectratio=true]{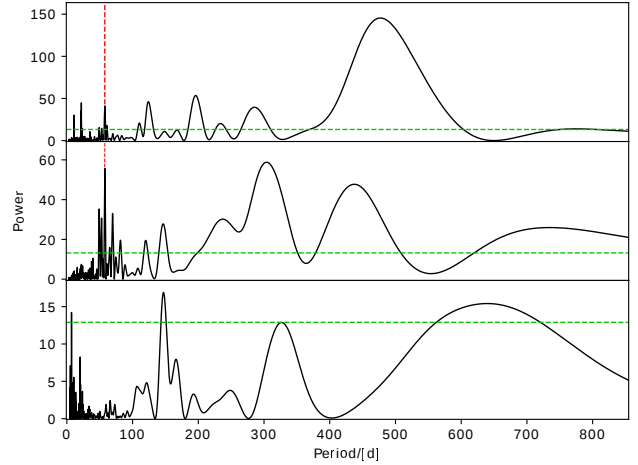}
  \caption{Period analysis of O-C data calculated from EB~A's primary eclipses (top panel) and secondary eclipses(middle panel) together with period spectrum obtained from EB~B's primary eclipses O-C data (bottom panel). Vertical red dashed line marks common period of 57\,d. Noise levels for the secondary times of minima from the EB~B eclipses were too high to resolve any peaks with sufficient confidence. Horizontal green dashed lines indicate power levels with false alarm probability of 0.1\,\%.} 
  \label{fig:periodogram}
\end{figure}

Accuracy of the minima times calculated in Sections~\ref{sec:EBAephemeris} and~\ref{sec:res_crv} enabled us to study eclipse time variations in more detail. After removing spikes caused by coincidence of EB~A and EB~B eclipses (see Section~\ref{sec:EBAephemeris}), we analyzed the remaining data using generalised Lomb-Scargle periodogram \citep{ZechmeisterKurster} which is a very effective method for analyzing unevenly spaced data. Horne Baliunas normalization was used and we searched for significant periods in range from Nyquist period up to 1400\,d. The same approach was used also for O-C data from EB B eclipses. Resulting period spectra are displayed in Fig. \ref{fig:periodogram}. 

\begin{figure}[t]
 \includegraphics[width=\columnwidth,keepaspectratio=true]{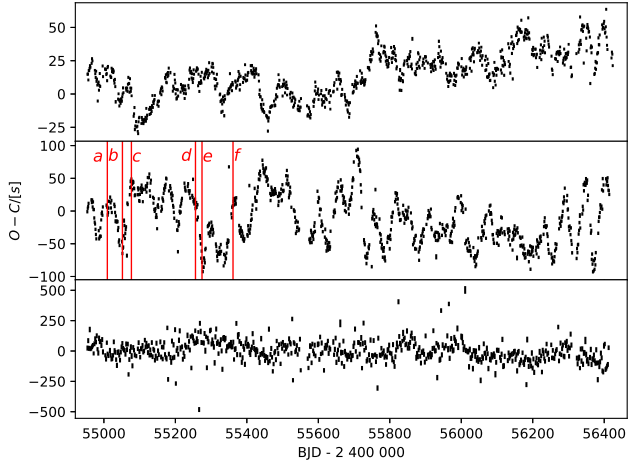}
  \caption{O-C diagram of primary and secondary eclipses of the EB~A with spikes removed (top and middle panel). Semi-regular variations with periods around 57\,d are clearly visible. Red vertical lines in the middle panel indicate the position of EB A's secondary eclipse residuals displayed in Fig. \ref{fig:spots}. Bottom panel shows O-C data from primary eclipses that belongs to the EB B eclipses where no significant period was detected. Due to very high levels of noise and lack of significant frequencies, O-C data from EB B's secondary eclipses were not displayed in this figure.}
  \label{fig:OC_fit}
\end{figure}

Periodic variations in O-C data can be a sign of the light-time effect (LITE) caused by the orbital motion of bodies already detected by eclipses or yet more bodies present in the system. However, filtered O-C data (Fig. \ref{fig:OC_fit}) and their corresponding period spectra in Fig. \ref{fig:periodogram} shows semi-regular behavior of the O-C data with no dominant period that would indicate presence of the LITE especially in EB A O-C data. 

Period of 57\,d was detected in the O-C data of the primary and secondary minima of EB~A (indicated by gray vertical dashed lines in Fig.~\ref{fig:periodogram}). However, no such period was detected in the data from EB~B. This fact combined with different amplitudes and phase shifts of the 57\,d O-C variations of primary and secondary minima are ruling out LITE as a possible explanation of this periodicity. Possible explanation for behavior of the O-C data is discussed in Section~\ref{sec:spots}.

\section{Presence of spots}
\label{sec:spots}

As we showed in the previous Section~\ref{sec:ETTV}, detected 57\,d periodicity of O-C variations cannot be explained by the gravitational interaction of both binaries nor by the presence of another body orbiting central eclipsing binary EB~A. Semi-regular nature of O-C variations of EB~A and its irregular amplitude changes, mainly for secondary minima, strongly suggest that variations in the O-C data can be caused by the presence of spot(s) on the surface of one or both components in the central EB~A.

We tested this hypothesis by subtracting template eclipse curve from observed EB~A's primary and secondary eclipses. For this purpose, template eclipse curve was obtained using the same approach as smoothed phase curve in section \ref{sec:phs_smth}, but during its creation, eclipses that produced spikes were omitted. Template eclipse function from section \ref{sec:EBAephemeris} was not suitable in this case because of artifacts present in the residual curve after subtracting the template eclipse function. Due to the insufficient sampling period of LC data, only eclipses covered by SC data were suitable for this analysis. 

\begin{figure}[t]
 \includegraphics[width=\columnwidth,keepaspectratio=true]{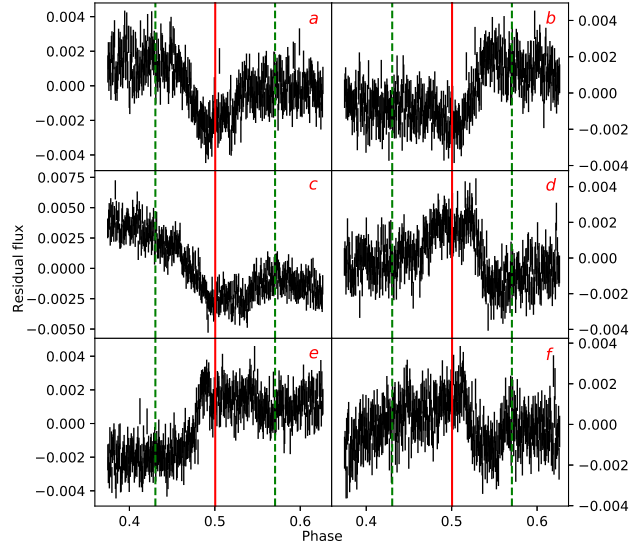}
  \caption{Set of 6 residuals curves of selected secondary eclipses of EB~A. Their positions in O-C diagram are indicated by vertical lines in Fig.~\ref{fig:OC_fit} designated by corresponding letter. Full red line in each sub-figure indicates the center of eclipse and pair of green dashed vertical lines shows start/end of the eclipses.}
  \label{fig:spots}
\end{figure}

Residuals of the EB A's primary eclipses showed no significant variations. On the other hand, residuals of secondary eclipses show noticeable variations as can be seen in Fig.~\ref{fig:spots}. We detected two types of variations in the residuals. The first type of variations (Fig.~\ref{fig:spots}~\textit{a},\textit{c},\textit{d},\textit{f}) is believed to be caused by the presence of a bright or dark spot on the surface of secondary component of EB~A. The second type of variations (Fig.~\ref{fig:spots}~\textit{b},\textit{e}) is of a similar nature but in this case flux does not return to its previous level. It is most likely due to the fact that the spot is no longer in the line of sight because of the rotation of the secondary component. Movement of these spots across the stellar surface or periodic occurrence of spots could be responsible for detected periods in O-C data of both primary and secondary eclipses. Presence of such spots on the surface of the EB A component periodically affects the overall shape of the light curve especially during eclipses and therefore the positions of primary or secondary minima can be shifted as it was observed.

\section{Discussion}
\label{sec:disscusion}

Although both sets of eclipses are clearly visible in the light curve of KIC~3832716, an overall hierarchy of the system is not obvious without spectroscopic observations.

Initially we analyzed the possibility of a hierarchic system with central eclipsing binary and circumbinary third component. We tried to determine the stability of such an orbit by numerical integration using Gauss-Radau integrator RA15 \citep{Everhart}. We were unable to achieve stable prograde orbit with orbital period of only 2.17 $d$. Periodic close encounters with components of central binary caused instability in such an arrangement of bodies. Although retrograde circular orbit was stable in the time range of our numerical integrations (150\,d), from the perspective of system's evolution however, such a tight retrograde orbit is very unlikely. More importantly, the dispersion of O-C data from second set of eclipses is too low to support the possibility of such an arrangement of components. The third body would in this situation eclipse primary or secondary component which is also moving around the system's center of mass. This would produce variations in O-C data with amplitude in the order of at least tens of minutes which was not observed (see fig. \ref{fig:OC_fit} bottom panel).

4-body configuration of the system is therefore the most likely explanation for these observations. In this configuration the system consists of two eclipsing binaries EB~A and EB~B orbiting around a common centre of mass. This arrangement of the components is much more consistent with observed data than the 3-body variant. The remaining question concerns the orbital period of the two EBs around a common center of mass. We analyzed the possibility that the detected period of 57\,d could be caused by LITE effect produced by the EB B or another body present in the system. However, different amplitudes of detected periods (by a factor of 4) suggest that those variations in O-C data are not caused by the LITE. Due to the absence of any other significant common periods in EB A period spectra we can conclude that EB B or other presumed body on high inclination orbit is not located in orbit around EB A with period lower than roughly 1400\,d (time span of Kepler's observations) because such an orbit would produce observable LITE effect in O-C data and the precision of detected minima times of the EB A is sufficient to discover such objects in its vicinity due to the LITE effect.

It leaves the possibility of EB B on long period orbit around EB A and also the possibility that these two EBs are not gravitationally bound at all. The first case is however more probable due to detected non-linear effects in O-C data from both of the EBs that we described by quadratic fit in ephemerides in Eq. \ref{eq:ephem} and Eq. \ref{eq:ephem2}. Their opposite effect on observed orbital periods of both EBs may be explained by the orbital motion of both EBs around common center of mass. Due to very short duration of observations compared to presumed long orbital period of the EBs we did not attempt to find parameters of such an orbit. Rough estimate of the mass ratio of EBs can however in this case be calculated by using relative rate of change of orbital periods:

\begin{equation}
 Q_{EB} = \frac{M_{EB_B}}{M_{EB_A}} = \frac{P_B \dot{P}_A}{P_A \dot{P}_B} = 0.7 \pm 0.3
\end{equation}
We can conclude that this estimation of mass ratio of EBs is consistent with luminosity and mass estimations of the components in Sec. \ref{sec:luminosity} assuming that both components of EB B are still on MS due to lower luminosity and insufficient time for them to evolve beyond MS.

\section{Conclusion}
We have identified that object KIC~3832716 consists of two eclipsing binaries with orbital periods $P_A\sim 1.1419$\,d and $P_B\sim 2.1703$\,d. that are in double EB configuration, orbiting a common center of mass with period much longer than the time span of Kepler photometric observations (1400\,d). From derived luminosities and effective temperatures we conclude that EB A system consists of MS primary component and a more evolved post MS secondary component. However, positions of the EB A components in the HR diagram (see Fig. \ref{fig:HR}) suggest more evolved but lighter secondary component that is impossible to explain by standard stellar evolution given the sensible assumption that those two components were formed at the same time from the same cloud of interstellar matter. This situation is very reminiscent of well known case of eclipsing binary Algol \citep{Baron12} which is a prototype for such behavior. The situation of "Algol paradox" was explained by mass transfer from secondary component after secondary component evolved beyond MS. We propose that the same mechanism is in action here as it can very well explain present state of the EB A, especially the relative positions of components in HR diagram and the fact that we consistently achieved very good quality of the light curve fit in case of the mass ratio between 0.5 and 0.6 for very wide range of inclinations which can be explained by unknown light contribution of the EB B system (see Fig. \ref{fig:qi_trial}). 

We also identified spot activity on the surface of secondary component, causing semi-regular variations in the observed times of minima of EB A with period about 57\,d. We presume that this period is connected with some sort of cycle in spot activity or it may be the period of longitudinal travel of long term spots across the surface of the secondary component of EB A system. 

EB B contains a red dwarf with luminosity below 0.09\,$L_{\odot}$ orbiting around primary component on circular orbit. Due to the large uncertainty in the light contribution of the EB B we were unable to derive mass ratio of both EBs but uncertainties of other parameters such as relative radii and effective temperatures or effective temperature ratio were affected very little by this issue.

Despite the fact that analyzed data from Kepler photometry are characterized with exceptional precision and duration, additional photometric and especially spectroscopic observations are required to confirm the findings of this paper. Further photometric observations of eclipses will help to confirm the gravitational bound between the EBs, their orbital period and hopefully other parameters of the orbit. Data from spectroscopic observations in combination with photometric observations could help to determine the absolute parameters of both EBs.

\section*{Acknowledgments}
This paper includes data collected by the Kepler mission. Funding for the Kepler mission is provided by the NASA Science Mission directorate. This work was supported by the Slovak Research and Development Agency under the contract No. APVV-15-0458. The research of M.F. was supported by the internal grant No. VVGS-PF-2018-758 of the Faculty of Science, P. J. \v{S}af\'{a}rik University in Ko\v{s}ice. The author would like to thank E. Paunzen and M. Lach for constructive criticism of the manuscript.

%% Non-BibTeX  (Name-Year style)
%


\begin{thebibliography}{}
% \bibitem[\protect\citeauthoryear{<author>}{<year>]{ref:?}
%    <ref. entry>
\bibitem[\protect\citeauthoryear{Armstrong et al.}{2014}]{Armstrong14}
Armstrong D. J., G\'{o}mez M. C., Y., Faedi F., Pollacco D., 2014, \mnras, 437, 4 
\bibitem[\protect\citeauthoryear{Baron et al.}{2012}]{Baron12}
Baron, F.; Monnier, J. D.; Pedretti, et al., 2012, \apj, 752, 20
\bibitem[\protect\citeauthoryear{Borucki et al.}{2010}]{Borucki}
Borucki W. J., Koch D., Basri G., et al., 2010, Science, 327, 977
\bibitem[\protect\citeauthoryear{Borucki et al.}{2011}]{Borucki11}
Borucki W. J., Koch D., Basri G., et al., 2011, \apj, 736, 1
\bibitem[\protect\citeauthoryear{Conroy et al.}{2014}]{Conroy}
Conroy E. K., Pr\v{s}a A., Stassun K. G., et al., 2014, \aj, 147, 45
\bibitem[\protect\citeauthoryear{Davenport}{2016}]{Davenport16}
Davenport J. R. A., 2016, \apj, 829, 23 
\bibitem[\protect\citeauthoryear{Everhart}{1985}]{Everhart}
Everhart E., 1985, in A. Carusi and G. B. Valsecchi, ed(s).,\textit{dynamics of comets: their origin and evolution, IAU Coll. No. 83}, Reidel, Dordrecht, The Netherlands, 185-202
\bibitem[\protect\citeauthoryear{Gies et al.}{2015}]{Gies}
Gies D.~E., Matson R.~A., Guo Z., et al., 2015, \aj, 150, 178
\bibitem[\protect\citeauthoryear{Gilliland et al.}{2010}]{Gilliland}
Gilliland R.~L., Jenkins J.~M., Borucki W.~J., et. al., 2010, \apjl, 713, 160
\bibitem[\protect\citeauthoryear{Howell et al.}{2014}]{Howell2014}
Howell S. B., et al., 2014, \pasp, 126, 398
\bibitem[\protect\citeauthoryear{Kurucz}{1993}]{Kurucz93}
Kurucz R. L., 1993, VizieR On-Line Data Catalogue VI, 39
\bibitem[\protect\citeauthoryear{Lindegren et al.}{2018}]{Lindegren18}
Lindegren L., Hern\'{a}ndez J., Bombrun A., et al., 2018,\\ https://doi.org/10.1051/0004-6361/201832727
\bibitem[\protect\citeauthoryear{Mikul\'{a}\v{s}ek}{2015}]{mikulasek}
Mikul\'{a}\v{s}ek Z., 2015, A\&A, 584, A8
\bibitem[\protect\citeauthoryear{Moln\'{a}r et al.}{2016}]{molnar16}
Moln\'{a}r L., Szab\'{o} R., Plachy E., 2016, Journal of the American Association of Variable Star Observers, 44, 168
\bibitem[\protect\citeauthoryear{Paxton et al.}{2018}]{Paxton18}
Paxton B., Schwab J., Bauer B. E., et al., 2018, \apjs, 234, 50
\bibitem[\protect\citeauthoryear{Pr\v{s}a}{2011a}]{Prsa}
Pr\v{s}a A., 2011a, PHOEBE scientific reference, Villanova university, p. 55
\bibitem[\protect\citeauthoryear{Pr\v{s}a et al.}{2011b}]{Prsa11}
Pr\v{s}a A., Batalha N., Slawson R. W., et al., 2011b, \aj, 141, 3
\bibitem[\protect\citeauthoryear{Pr\v{s}a \& Zwitter}{2005}]{Prsa_Zwitter}
Pr\v{s}a A., {Zwitter} T., 2005, \apj, 628, 426
\bibitem[\protect\astroncite{Sartoretti et al.}{2018}]{Sartoretti18}
Sartoretti P., Katz D., Cropper M., el al., 2018, \aap, 616, A6
\bibitem[\protect\astroncite{Schlafly \& Finkbeiner}{2011}]{Schlafly11}
Schlafly, E.F., Finkbeiner, D.P.  2011, \apj, 737, 103
\bibitem[\protect\astroncite{Skrutskie et al.}{2006}]{Skrutskie06}
Skrutskie M. F., Cutri R. M., Stiening R. et al., 2006, \aj, 131, 2
\bibitem[\protect\astroncite{Slawson et al.}{2011}]{Slawson11}
Slawson R., Pr\v{s}a A., Welsh W. F., et al., 2011, \aj, 142, 5
\bibitem[\protect\citeauthoryear{Stellingwerf}{1978}]{Stellingwerf}
Stellingwerf R. F., 1978, \apj, 224, 953
\bibitem[\protect\citeauthoryear{van Hamme}{1993}]{vanHamme}
van Hamme W., 1993, \aj, 106, 2096
\bibitem[\protect\citeauthoryear{Zasche et al.}{2015}]{Zasche}
Zasche P., Wolf M., Ku\v{c}\'{a}kov\'{a} H., 2015, \\ preprint (arXiv:1506.00848v1)
\bibitem[\protect\citeauthoryear{Zechmeister \& K\"{u}rster}{2009}]{ZechmeisterKurster}
Zechmeister M., K\"{u}rster M., 2009, \aap, 496, 577

% \bibitem[\protect\citeauthoryear{<author>}{<year>]{ref:?}
%    <ref. entry>
\end{thebibliography}
\end{document}